\documentclass[%
 reprint,
 amsmath,amssymb,
 aps, 
prb,
]{revtex4-1}

\usepackage{graphicx}
\usepackage{dcolumn}
\usepackage{bm}

\usepackage{hyperref}
\usepackage[capitalise]{cleveref}
\usepackage{nameref}

\begin{document}

\preprint{APS/123-QED}

\title{A Bayesian approach to RFI mitigation}
\author{S. A. K Leeney}
 \altaffiliation[Also at ]{Kavli Institute for Cosmology, Madingley Road, Cambridge, CB3 0HA, UK}
 \email{sakl2@cam.ac.uk} 
\author{W. J Handley }%
 \email{wh260@cam.ac.uk} 
 \altaffiliation[Also at ]{Kavli Institute for Cosmology, Madingley Road, Cambridge, CB3 0HA, UK}
\author{E. de Lera Acedo}%
 \email{ed330@cam.ac.uk}
 \altaffiliation[Also at ]{Kavli Institute for Cosmology, Madingley Road, Cambridge, CB3 0HA, UK}
\affiliation{%
 The University of Cambridge, Astrophysics Group, Cavendish Laboratory, J. J. Thomson Avenue, Cambridge, CB3 0HE, UK} 

\date{\today}

\begin{abstract}
Interfering signals such as Radio Frequency Interference from ubiquitous satellite constellations are becoming an endemic problem in fields involving physical observations of the electromagnetic spectrum. To address this we propose a novel data cleaning methodology. Contamination is simultaneously flagged and managed at the likelihood level. It is modeled in a Bayesian fashion through a piecewise likelihood that is constrained by a Bernoulli prior distribution. The techniques described in this paper can be implemented with just a few lines of code. 
\end{abstract}

\maketitle

\section{Introduction}
Satellite constellations in low earth orbit such as SpaceX's Starlink will likely number 100,000 by 2030~\cite{venkatesan2020impact}. Described in a recent Nature article as `horrifying'~\cite{witze2022satellite}, the impact of Radio Frequency Interference (RFI) created by these `megaconstellations' on Astronomy is of significant concern. Interfering signals like RFI are problematic because they cause information in contaminated frequency channels to be lost which can lead to significant systematic error if not properly modelled.

The frequency bands that these satellites and other modern telecommunications devices operate in are not protected or reserved for Astronomy. They clash with current and next generation experiments such as the ngVLA~\cite{mckinnon2019ngvla} and the SKA~\cite{bourke2015advancing}, which plan probe yet unseen epochs of the universe. Furthermore, the latest telescopes operate in large regions of the sky and across wide frequency bands, making it impossible to avoid these satellite swarms. This is pushing Astronomy to remote corners of earth and beyond, to space~\cite{sabelhaus2004overview}.

However, moving projects to space is not a long term solution. There is only a short time window until the RFI-quiet dark side of the moon becomes contaminated by satellites and other devices required by projects such as LuSEE\cite{giardino2019impact} and the LCRT~\cite{bandyopadhyay2021lunar} which themselves are hoping to exploit the clear lunar skies. Furthermore space itself is filled with cosmic rays, which lead to RFI-like interference and thus have been problematic for the JWST's near infrared spectrometer~\cite{giardino2019impact}.

Consequentially, RFI is becoming the fundamental bottleneck in modern Astronomy and beyond; with modern observations being at the forefront of many fields of fundamental Physics, such as dark matter detection. New statistical techniques are therefore urgently required to address this rapidly growing problem.

Furthermore, projects such as the Event Horizon Telescope~\cite{event2019first}, Planck~\cite{ade2014planck} and BICEP~\cite{ade2021bicep} all used Bayesian statistics in their data analysis pipelines, but there is currently no way to manage RFI in a Bayesian sense forcing astronomers to patch traditional RFI mitigation algorithms into their Bayesian systems. Modern projects implement Singular Value Decomposition~\cite{offringa2010post} watershed segmentation~\cite{kerrigan2019optimizing}, Deep Learning methods~\cite{akeret2017radio};~\cite{vafaei2020deep};~\cite{sun2022robust} and more recently Gibbs sampling~\cite{2022arXiv221105088K}.

Scientists are constantly searching for more elusive and faint signals, thus data cleaning is becoming increasingly important. The Laser Interferometer Gravitational-wave Observatory (LIGO)~\cite{abbott2009ligo}, for example, is sensitive enough to detect a car starting miles away and thus is extremely sensitive to data corruption~\cite{ormiston2020noise}. The next generation aLIGO~\cite{harry2010advanced} will be even more so. As such there is a need for novel data cleaning methodologies beyond fields involving measurements of the electromagnetic spectrum.

In this paper we propose a data cleaning methodology that takes a Bayesian approach, where contamination is both flagged and managed at the likelihood level. Detailed benchmarking of this technique will be provided when it is applied on data from a next generation low frequency global experiment, REACH~\cite{de2022reach}, which will see first light this year. A usage example of the methodologies described in this work can be found at \href{https://github.com/samleeney/Publications}{github.com/samleeney/publications}.

\section{Theory}
\subsection{Bayesian Inference}
\label{sec:bayesianinftheory}
Bayesian methods can be used to perform parameter estimates and model comparison. A model $\mathcal{M}$ uses data $\mathcal{D}$ to infer its free parameters $\theta$. Using Bayes Theorem,
\begin{align}
    P(\mathcal{D}|\theta) \times P(\theta) &= P(\theta|\mathcal{D}) \times P(\mathcal{D}), \\
    \mathcal{L} \times \pi &= \mathcal{P} \times \mathcal{Z},
\end{align}
the prior $\pi$ is updated onto the posterior $\mathcal{P}$ in light of the likelihood $\mathcal{L}$ and furthermore the Bayesian Evidence $\mathcal{Z}$ can be inferred by computing the integral
\begin{equation}
    \mathcal{Z} = \int \mathcal{L}(\theta) \times \pi(\theta) \; d\theta.
\end{equation}
In practice, $\mathcal{P}$ and $\mathcal{Z}$ can be determined simultaneously using a Bayesian numerical solver. We use the Nested Sampling algorithm \textsc{polychord}~\cite{handley2015polychord}; where a series of points generated within $\pi$ are updated such that they sequentially contract around the peak(s) of the likelihood, forming the posterior which can be used to generate parameter estimates. The artefacts of this process can then be used to compute $\mathcal{Z}$, which is used for model comparison. For a more detailed description of Bayesian Inference and Nested Sampling see~\cite{mackay2003information}~\cite{sivia2006data}.

\subsection{Data cleaning likelihood}\label{sec:rficorrtheory}
A sufficiently contaminated data point can be considered corrupted. Any information relevant to the model is lost and furthermore it cannot be modelled as Gaussian noise. Assuming $\mathcal{D}$ is uncorrelated, the likelihood
\begin{equation}
\begin{aligned}
   \mathcal{L} &= P(\mathcal{D}|\theta) = \prod_{i} \mathcal{L}_{i}(\theta),\label{l1} = \prod_{i} P(\mathcal{D}_i|\theta)
\end{aligned}
\end{equation}
where $i$ represents the $i$'th data point, is insufficient to model such contaminated data.
It is therefore necessary to model the likelihood that each data point is corrupted. 

Thus, we introduce a piecewise likelihood including the possibility of corruption of data
\begin{equation}
    P(\mathcal{D}_i|\theta) = \begin{cases}
        \mathcal{L}_i(\theta) &: \text{uncorrupted}\\
        \Delta^{-1}[ 0<\mathcal{D}_i<\Delta] &: \text{corrupted}.\\
    \end{cases}
\end{equation}
Corruption is modelled as the data becoming completely unreliable and therefore being distributed uniformly within some range $\Delta$ (which, as a scale of corruption, has the same dimensions as the data).
An efficient way to write this likelihood is
\begin{equation}
    P(\mathcal{D}|\theta, \varepsilon) = \prod_{i} \mathcal{L}_{i}^{\varepsilon_{i}} \Delta^{\varepsilon_i-1}
    \label{eq:li2}
\end{equation}
where the Boolean mask vector $\varepsilon$ has a $i$th component which takes the value $1$ if the datum $i$ is uncorrupted and value $0$ if corrupted.

We do not know before the data arrive whether or not they are corrupted. We may infer this in a Bayesian fashion, by ascribing a Bernoulli prior probability $p_i$ of corruption (which has dimensions of probability) i.e:
\begin{equation}
P(\varepsilon_i) = p_i^{(1-\varepsilon_i)}(1-p_i)^{\varepsilon_i}.\label{eq:pei}
\end{equation}
Both $\Delta$ and $p_i$ are required for a dimensionally consistent analysis. It should be noted that above we assume the a-priori probability that each bin is corrupted is uncorrelated, i.e $P(\varepsilon)=\prod _i P(\varepsilon_i)$, which in practice will almost certainly not be true. We will discuss later the extent to which this assumption can be considered valid.

Multiplying \cref{eq:pei,eq:li2} yields
\begin{equation}
    P(\mathcal{D},\varepsilon|\theta) = \prod_{i} \left[\mathcal{L}_{i}(1-p_i)\right]^{\varepsilon_{i}} \left[p_i/\Delta\right]^{(1-\varepsilon_i)}
    \label{eqn:likelihood_eps}
\end{equation}
and to recover a likelihood independent of $\varepsilon$ we formally can marginalise out:
\begin{align}
    P(\mathcal{D}|\theta) &=\sum_{\varepsilon \in \{ 0, 1 \} ^N}P(\mathcal{D},\varepsilon|\theta) \\
    &= \sum_{\varepsilon \in \{ 0, 1 \} ^N} \prod_{i} \left[\mathcal{L}_{i}(1-p_i)\right]^{\varepsilon_{i}} \left[p_i/\Delta_i\right]^{(1-\varepsilon_i)}.
    \label{eq:pdtheta2}
\end{align}
This would require the computation of the all $2^N$ terms in \cref{eq:pdtheta2}. For realistic values of $N$, this computation becomes impractical. However, if it is assumed that the most likely model (i.e the maximum term in \cref{eq:pdtheta2}) dominates over the next to leading order terms, we can make the approximation
\begin{equation}
    P(\mathcal{D},\varepsilon|\theta) \approx \delta_{\varepsilon \varepsilon^\mathrm{max}} \times P(\mathcal{D},\varepsilon^{\mathrm{max}}|\theta)\label{eqn:postierioreps}
\end{equation}
where $\delta_{ij}$ is the usual Kroneker delta function, and
 $\varepsilon^\mathrm{max}$ is the mask vector which maximises the likelihood $P(D,\varepsilon|\theta)$, namely:
\begin{equation}
    \varepsilon^{\mathrm{max}}_{i}=
    \begin{cases}
        1, & \mathcal{L}_i(1-p_i) > p_i/\Delta_i \\
        0, & \text{otherwise}.
    \end{cases}
    \label{eqn:emax}
\end{equation}
Under this approximation we find that the sum in \cref{eq:pdtheta2} becomes
\begin{equation}
    P(\mathcal{D}|\theta) \approx P(\mathcal{D},\varepsilon^{\mathrm{max}}|\theta).\label{eq:approx}
\end{equation}

In practice the approximation in \cref{eq:approx} is only valid if the next to leading order term is much smaller, such that
 \begin{equation}
 P(\mathcal{D}|\theta, \varepsilon_{\mathrm{max}}) \gg \mathrm{max}_j P(\mathcal{D}|\theta,\varepsilon^{(j)})\label{eq:nlo},
\end{equation}
where $\varepsilon^{(j)}$ is $\varepsilon^\mathrm{max}$ with its $j$th bit flipped:
\begin{equation}
    \varepsilon^{(j)}_k = 
    \begin{cases}
    1-\varepsilon^{\mathrm{max}}_k & k=j \\
    \varepsilon^{\mathrm{max}}_k & k\ne j \\
\end{cases}
\end{equation}
and we may use \cref{eq:nlo} as a consistency check.

To summarise, we can correct for contamination under these approximations by replacing the original likelihood $\mathcal{L} = \prod_i\mathcal{L}_i$ in \cref{l1} with 
\begin{equation}
    P(\mathcal{D}|\theta) = \prod_i\left[\mathcal{L}_{i}(1-p_i)\right]^{\varepsilon^{\mathrm{max}}_{i}} \left[p_i/\Delta\right]^{(1-\varepsilon^\mathrm{max}_i)}
    \label{eqn:likelihood}
\end{equation}
where $\varepsilon^{\mathrm{max}}$ is defined by \cref{eqn:emax}.
\subsection{Computing the posterior}
The posterior and evidence are computed from \cref{eqn:likelihood} via Nested Sampling (although any numerical Bayesian sampling method could be used). Taking logs for convenience gives
\begin{equation}
    \begin{aligned}
        \log{P(\mathcal{D}|\theta)} = \sum_{i} &[\log{\mathcal{L}_i}+\log({1-p_i})]\varepsilon^{\mathrm{max}}\\ 
        &+ [\log{p}_i - \log{\Delta}](1 - \varepsilon^\mathrm{max}_i), \label{eq:loglikelihood}
    \end{aligned}
\end{equation}
yielding a masked chi squared like term which can be used to distinguish whether there is a statistically significant difference between the classes of data, i.e corrupted or non corrupted. Furthermore, the second term in \cref{eq:loglikelihood} introduces an Occam penalty. Each time a data point is predicted to be contaminated its likelihood is replaced with the penalty rather than being completely removed. Without this term, the likelihood where all data points are flagged would be larger and thus `more likely' than all other possibilities. Therefore, flagging all datum would always be preferable.

We compute this by imposing the condition in \cref{eqn:emax} on \cref{eq:loglikelihood} as follows, \begin{equation}
    \log{P(\mathcal{D}|\theta)} =  
    \begin{cases}
        \log \mathcal{L}_i + \log (1-p_i), &
    \begin{aligned}
        &[\log{\mathcal{L}_i} + \log({1-p_i}) \\
        &> \log p_i - \log \Delta]
    \end{aligned}\\ 
        \log p_i - \log \Delta, & \text{otherwise}.
    \end{cases}
    \label{eqn:loglcompute}
\end{equation}
The corrected likelihood is then updated iteratively via the selected Bayesian sampling method, compressing the prior onto the posterior while simultaneously correcting for conamination. One may also notice that the condition $\log \mathcal{L}_i + \log(1-p_i) > \log p_i - \log \Delta$ in~\cref{eqn:loglcompute} relates to a Logit function, such that
\begin{equation}
\log \mathcal{L}_i + \log \Delta  > \textsc{logit(p)}.   
\end{equation}
Logit functions are used routinely as an activation function in binary classification tasks, hinting at the potential of a future extension of this work using machine learning.

\section{Toy example}\label{sec:toyex}
We will initially test this approach on a simple toy model with a basic contaminant signal injected. We then move onto a more realistic and complex case in \cref{sec:reachex}.

\subsection{Initial Testing}\label{sec:initialsetup}
Two simple datasets are generated for comparison consisting of a line with $m=1$ and $c=1$ with $\sigma=5$ order Gaussian noise. One is contaminated by an RFI like signal and the other (the ground truth) is not. They are fit in a Bayesian sense, attempting to recover the two free parameters $m$ and $c$ using the correcting likelihood in \cref{eqn:loglcompute} with
\begin{equation}
\mathcal{L}_i = -\frac{\log(2\pi \sigma^{2})}{2} - \frac{[y_{i} - y_{\mathcal{S}}(x_i;m, c)]^2}{2\sigma^2},
\end{equation}
where $\theta = m, c, \sigma$, $y_i$ is the simulated data and $y_\mathcal{S}(x_i; m, c)$ are the parameter estimates at the $i$'th sampling iteration are used to compute the model $y_i=mx_i + c$. We set $\Delta = \mathcal{D}_\textsc{max}$, to encapsulate the full range of possible data values. $\Delta$ could likely be fit as a free parameter, as will be investigated further in future works.

Evaluating the posterior on $\varepsilon$ we can assess how frequently across the entire sampling run each datum was believed to fit (non corrupted) or not fit (corrupted) the model. Contaminated points should make up a near zero fraction of the posterior. Conversely, points that are not contaminated would often fit the model and as such contribute significantly to the final posterior distribution. There can also be some points that lie somewhere in between, which the model is less confident are uncontaminated. This may occur with datum that deviate the most from the true signal due to higher order Gaussian noise, for example.

It should be emphasised that although $\varepsilon_i$ is constrained to binary values, the subsequent mask on $\varepsilon$ is not. Unlike traditional RFI flagging algorithms, points are not classified in a binary manor. The mask takes the weighted mean across the posterior. Thus, points more likely to contain RFI will have less `impact' on the final posterior distribution than points believed to be uncontaminated. The mask could be thought of as being slightly opaque to these data points, accounting for the models uncertainty. Incorporating the models confidence in its correction directly into the subsequent parameter estimates makes this approach unique in comparison with its counterparts.

\subsection{Model Evaluation}
\begin{figure*}
	\includegraphics[width=\textwidth]{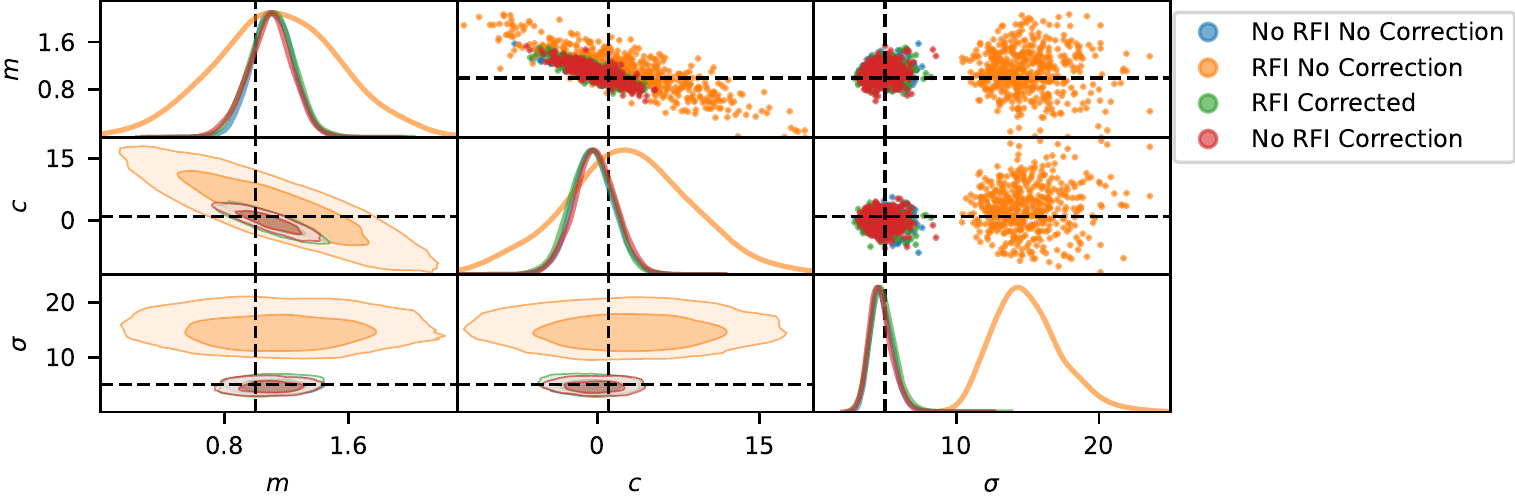}
    \caption{Showing the parameter distributions inferred from the dataset described in~\cref{sec:initialsetup}. The top left to bottom right panes show probability distribution functions for $m$, $c$ and $\sigma$, respectively. Plots generated using  posterior plotting tool \textsc{anesthetic}~\protect\cite{anesthetic}.}
    \label{fig:anesthetic}
\end{figure*}

The two aforementioned data sets are evaluated when fit using the likelihood capable of correcting for contamination and also when using a standard likelihood, which cannot natively account for contamination. This generates a total of four posterior distributions for comparison. Of these, all but the contaminated, uncorrected case would be expected to perform similarly if RFI has been effectively mitigated. From a Bayesian standpoint, the simplest model will always be preferable. Thus, for the clean dataset it would be expected that the standard likelihood would be preferred slightly over the correcting likelihood. Fig.~\ref{fig:anesthetic} shows the parameter distributions inferred from the data. The results are as expected; parameters in all but the `RFI No correction' are inferred to within $1\sigma$ of their true value.

\begin{figure}
	\includegraphics[width=\columnwidth]{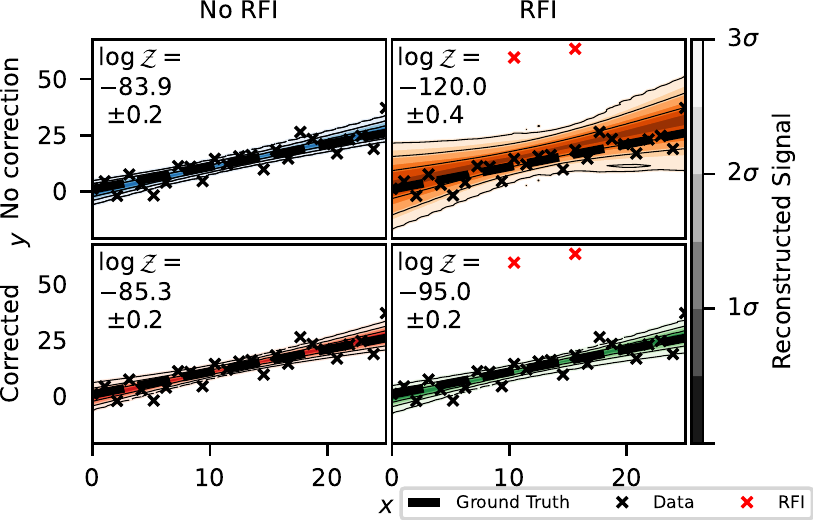}  
    \caption{Showing the inferred parameter estimates in a contour plot, where darker tones indicate higher $\sigma$ confidence in the parameter estimates. Generated from the dataset described in Section~\ref{sec:initialsetup}. The Bayes factor is $-1.4\pm0.3$ for the no RFI case and $25.0\pm0.4$ for the RFI case. The plots are generated using the functional posterior plotter \textsc{fgivenx}~\protect\cite{fgivenx}.}
    \label{fig:fgx}
\end{figure}

As seen in~\cref{fig:fgx}, the model that does not correct for RFI is slightly preferred for uncontaminated data. Conversely the correcting model is strongly preferred on the contaminated data indicating that the correction is working as predicted.

Viewing the posterior plots of $P(y|x, \mathcal{D})$ in \cref{fig:fgx} it is clear that when RFI is not corrected, the true parameter values are outside the $1\sigma$ and sometimes $2\sigma$ confidence bounds. Conversely the other three cases fit almost entirely within the $1\sigma$ bounds, indicating that the RFI has been mitigated.

\subsection{Evaluating the $\log p$ dependence}\label{sec:logpdependence}
Proper selection of the probability thresholding term $\log p$ is essential. From a Bayesian standpoint it should be set to represent our prior degree of belief in there being RFI in each datum. We assess the $\log p$ dependence while varying $\log p$ as a function of the RMSE on the fit generated from the parameter estimates, the $\log$ Bayesian Evidence and the mean number of points flagged across all samples.

\begin{figure}
    \centering
	\includegraphics[width=\columnwidth]{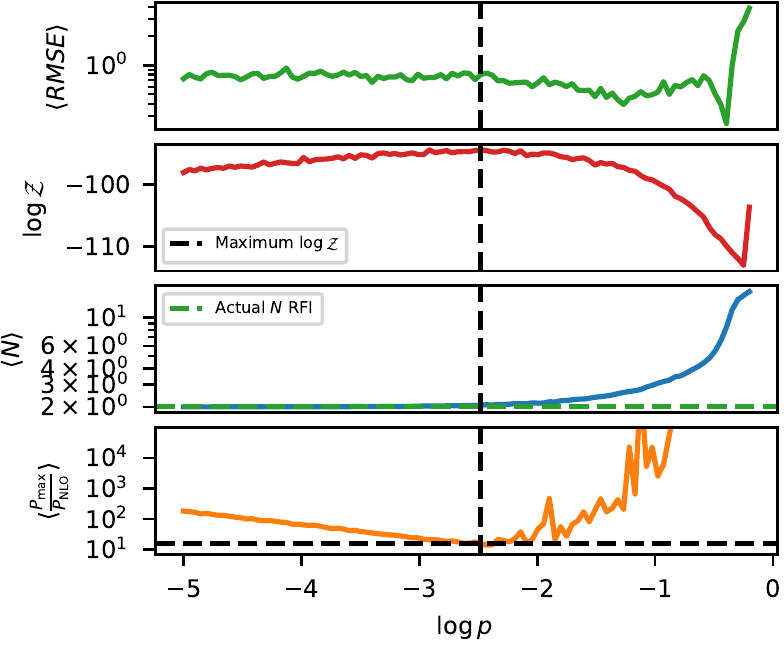}
    \caption{Assessing how various methods of model evaluation vary as function of $\log p$. From top to bottom: the RMSE, the $\log$ of the Bayesian Evidence, then the weighted average number of points flagged and finally the radio between $P_{\textsc{max}}$ and $P_{\textsc{NLO}}$. All dependant variables (excluding $\log \mathcal{Z}$) are averaged over the weighted posterior samples. The noise observable in these plots is sampling noise; the noise in the simulated data is seeded.}
    \label{fig:4pane}
\end{figure}
For high $\log p$, the RMSE is high and we observe in \cref{fig:4pane} that the model generates less accurate parameter estimates. Here the threshold is so high that the model is more confident that any of the points are RFI than non RFI. This matches the corresponding low evidence. The RMSE drops as $\log p$ decreases to near its minimum. The model incorrectly flags $\approx 5$ data points while the RMSE is low, showing that the model is able to generate accurate parameter estimates whilst over flagging, indicating it is insensitive to false positives. As $\log p$ decreases further, the model is better able to distinguish between higher order Gaussian noise and as such the average number of points predicted to be RFI approaches the true value. As this happens the evidence also reaches its maximum, which indicates that the Bayesian Evidence is appropriately showing how well each of the many models created by different $\log p$ values fit the data.

\subsection{To what extent is $P(\mathcal{D}|\theta) \approx P(\mathcal{D}|\theta, \varepsilon_{\mathrm{max}})$ valid?}
A key assumption is made in \cref{eq:approx} is that the leading order term, \cref{eq:loglikelihood}, is considerably larger than all the other possible terms for $\varepsilon \in (0, 1)^N$. It is necessary to test the validity of this approximation by computing \cref{eqn:loglcompute} and comparing the result ($P_{\textsc{max}}$) with the next leading order term ($P_{\textsc{NLO}}$) as calculated by \cref{eq:nlo}. For $-5 < \log p < -0.1$. These results are displayed in the bottom pane of \cref{fig:4pane}. $P_{\textsc{max}}$ is 18 times larger than $P_{\textsc{NLO}}$ at peak $\log \mathcal{Z}$ and increases linearly for $\log p$ below this. Depending on the $\log p$ selection strategy, $P_{\textsc{max}}$ is at least 11 times more likely than the next leading order term. Assuming an appropriate $\log p$ selection strategy, the ratio would be higher, indicating that $P(\mathcal{D}|\theta) \approx P(\mathcal{D}|\theta \varepsilon_{\mathrm{max}})$ is valid.

\subsection{Selection Strategy for $\log p$}
Various selection strategies could be taken to select the optimal $\log p$ value. For each $\log p$, the model changes. As such, selecting the $\log p$ that maximises the evidence seems to be the most obvious selection strategy. In the case of the toy model, the peak $\log \mathcal{Z}$ occurs where $\log p = -2.7$. Here, $P_{\textsc{max}}$ is 18 times larger than $P_{\textsc{NLO}}$. Another possible strategy could be to select $\log p$ where the number of points flagged is at its minimum. 

It is also possible to ascribe a prior to $\log p$, fitting it as a free parameter thus fully automating the approach. This will be examined further in future works.

\section{REACH Example}\label{sec:reachex}
\begin{figure}
	\includegraphics[width=\columnwidth]{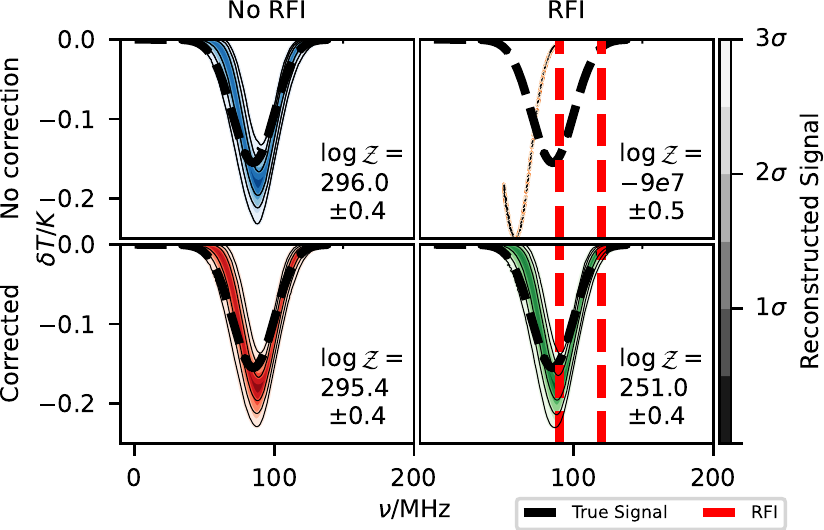}
    \caption{Showing the results when the RFI correction is applied on simulated data in the REACH data analysis pipeline. The Bayes factor for the no RFI case is $-0.6\pm0.6$ and the Bayes factor for the RFI case is $9e7\pm0.6$.}\label{fig:reach_dual_plot}
\end{figure}
Finally, we examine a real use case for this method. The REACH~\cite{de2022reach} radio telescope is designed to detect the 21cm signal from the Cosmic Dawn. We select REACH as a testing ground for our methodology because it operates in the same low frequency ranges as the next generation of Astronomical experiments, such as the nvGLA the SKA and takes a Bayesian approach to data analysis~\cite{anstey2021general}. The 21cm signal is expected to take the shape of an inverted Gaussian, so the model takes the form
\begin{equation}
    f(x) = A \exp{\bigg( -\frac{(x-\mu)^{2}}{2 \sigma^{2}}\bigg)}
\end{equation}
with center frequency $\mu$, standard deviation $\sigma$ and magnitude $A$ all free parameters. The four cases discussed in \cref{sec:toyex} are then examined, but this time on a simulated sky data set containing a 21cm signal with two RFI spikes injected.
The No RFI Correction and ground truth cases are very similar with the simpler (ground truth) case marginally preferred as expected. The RFI Corrected case is again similar with a slightly lower evidence due to the penalties incurred during the corrections. The above is also evident when viewing \cref{fig:reach_dual_plot}. The reconstructed signal is within $2\sigma$ of the true signal for all but the RFI No Correction case.

\section{Conclusions}\label{sec:conclusions}
In this paper, which serves as a proof-of-concept, we show that contamination can be both cleaned and corrected in a fully Bayesian sense at the likelihood level. We demonstrate our general approach in the context of signal processing for Astronomy, but these methods will likely be useful beyond. Forthcoming results from current state of the art low frequency Astronomy experiments~\cite{de2019reach} (where these methods will be benchmarked in the coming months) and extensive testing of this approach as a general Bayesian data cleaning methodology will be used to provide a more detailed analysis in the future.

\nocite{*}

\bibliography{ref}

\end{document}